\documentclass[sigplan,10pt]{acmart}
\renewcommand\footnotetextcopyrightpermission[1]{}
\AtBeginDocument{%
  }

\setcopyright{acmlicensed}
\copyrightyear{2018}
\acmYear{2018}
\acmDOI{XXXXXXX.XXXXXXX}
\acmISBN{978-1-4503-XXXX-X/2018/06}

  \usepackage{enumitem}
\usepackage{xcolor, colortbl, booktabs}
\usepackage{algorithm}
\usepackage{algorithmic}
\usepackage{comment}
\usepackage{makecell}
\usepackage{array}
\usepackage{xspace}
\newcommand{\name}{\textit{RelayMoE}\xspace}

\definecolor{jggreen}{RGB}{0,120,60}

\begin{document}


\title[]{Memory-Efficient Expert Routing for Distributed MoE Training}



\author{Arnab Kanti Tarafder}
\email{aktarafder@wm.edu}
\affiliation{%
  \institution{William \& Mary}
  \city{}
  \state{}
  \country{}
}

\author{Jaume Guasch-Martí}
\email{jaume.guasch@bsc.es}
\affiliation{%
  \institution{Barcelona Supercomputing Center}
  \city{}
  \country{}
  }

\author{Gokcen Kestor}
\email{gokcen.kestor@bsc.es}
\affiliation{%
  \institution{Barcelona Supercomputing Center}
  \city{}
  \country{}
  }

\author{Jie Ren}
\email{jren03@wm.edu}
\affiliation{%
  \institution{William \& Mary}
  \city{}
  \country{}
  }

\renewcommand{\shortauthors}{Paper \#2251}

\begin{abstract}
As Mixture-of-Experts (MoE) models scale toward hundreds of experts and higher top-k routing, memory efficiency in distributed training becomes a critical bottleneck. Peak memory is dominated by the MoE block, not attention: every intermediate buffer in the MoE dispatch pipeline is individually scaled by top-k routing. The standard all-to-all dispatcher sends all routed tokens in a single collective step, requiring the full top-k-expanded buffer to be constructed at once.


In this work, we propose \name, a ring-based MoE execution model that computes locally as expert weights or tokens circulate, avoiding full top-$k$-expanded dispatch buffers. \name selects between expert and token routing according to communication volume and overlaps transfers with computation. The ring structure naturally supports memory-efficient MoE recomputation during backward: each hop reconstructs expert intermediates, uses them to compute gradients, and releases them before the next hop. The saved memory supports longer sequences and larger batches, or retains more attention activations to reduce attention recomputation and improve training throughput. We evaluate \name on 30B--57B production MoE models and varied expert configurations. In single-layer MoE experiments, \name achieves a 2$\times$ average speedup over Megatron-LM. In full-model training under the same GPU memory budget, it improves throughput by up to 2.02$\times$ and extends the largest tested trainable sequence length by up to 2.85$\times$.

\end{abstract}

\settopmatter{printacmref=false}
\renewcommand\footnotetextcopyrightpermission[1]{}
\pagestyle{plain}

\settopmatter{printfolios=true}

\maketitle

\section{Introduction}

The rapid scaling of large language models (LLMs) has increasingly relied on Mixture-of-Experts (MoE) architectures. By activating only a subset of experts per token, MoE models achieve stronger quality at a fraction of the compute cost of dense models~\cite{rajbhandari2022deepspeedmoe,fedus2022switch, du2022glam}, making them a common architecture for frontier systems. This advantage has led to larger and sparser MoE configurations along multiple dimensions simultaneously: more experts, higher top-k routing, longer training sequences, and wider expert Feedforward Network (FFN) layers, as shown in Table~\ref{tab:trend}. Each of these dimensions independently improves model quality, but together they increase GPU memory requirements during distributed training.

\begin{table}[t]
\centering
\caption{MoE architecture trends in frontier models. Buffer Multiplier shows how each dispatch buffer is scaled relative to the base activation tensor by top-$k$ routing. Combined with growing sequence lengths (32K$\to$262K), peak memory from MoE dispatch has grown by an order of magnitude. 
}
\label{tab:trend}
\scriptsize
\begin{tabular}{lcccccc}
\toprule
\multicolumn{1}{l}{\textbf{Model}} & \textbf{Date} & \makecell{\textbf{\# of} \\ \textbf{\tiny{Experts}}} & \makecell{\textbf{\tiny{Top-}} \\ \textbf{\textit{k}}} & \makecell{\textbf{\tiny{Expert}} \\ \textbf{\tiny{FFN}}} & \makecell{\textbf{Max} \\ \textbf{\tiny{Context}}} & \makecell{\textbf{\tiny{Buffer}} \\ \textbf{\tiny{Mult.}}} \\
\midrule
\tiny{Mixtral 8x22B \cite{jiang2024mixtral}}   & 04/24 & 8   & 2  & 16384 & 32K  & 2$\times$ \\
\tiny{DeepSeek-V3 \cite{deepseek2024v3}}       & 12/24 & 256 & 8  & 2048  & 128K & 8$\times$ \\
\tiny{Qwen3-30B-A3B \cite{yang2025qwen3technicalreport}}        & 04/25 & 128 & 8  & 768   & 131K & 8$\times$ \\
\tiny{Kimi K2 \cite{kimiteam2026kimik2openagentic}}            & 07/25 & 384 & 8  & 2048  & 128K & 8$\times$ \\
\tiny{Qwen3-Next-80B-A3B \cite{qwen2025qwen3next}}    & 09/25 & 512 & 10 & 512   & 128K & 10$\times$ \\
\tiny{Qwen3.5-35B-A3B \cite{qwen2026qwen35}}   & 02/26 & 256 & 8  & 512   & 262K & 8$\times$ \\
\tiny{Qwen3.5-397B-A17B \cite{qwen2026qwen35}}   & 02/26 & 256 & 8  & 512   & 131K & 8$\times$ \\

\bottomrule
\end{tabular}
\vspace{-10pt}
\end{table}

To train these increasingly large MoE models, expert parallelism (EP)~\cite{shazeer2017outrageously, rajbhandari2022deepspeedmoe} distributes experts across GPUs and dispatches tokens to their assigned experts via all-to-all collective communication. However, the all-to-all dispatcher requires the full top-k-expanded buffer to be constructed before communication can begin. Because each token is replicated to its top-k selected experts, every intermediate buffer in the dispatch pipeline is $k\times$ larger than the base activation tensor. This compounds further with increasing sequence lengths and batch sizes.
As a result, peak memory from MoE dispatch has grown by an order of magnitude in recent architectures (Table~\ref{tab:trend}). Meanwhile, attention memory has been effectively reduced through memory-efficient algorithms~\cite{flashattn2022, flashattn2_2024, ringattention2024, ulysses2023, korthikanti2023reducing} and hybrid linear-attention designs in recent architectures~\cite{qwen2026qwen35}, leaving MoE dispatch as the dominant contributor to peak memory. As Figure~\ref{fig:m1} shows, MoE blocks create transient memory spikes that determine the OOM boundary and limit the maximum trainable sequence length and batch size.

Prior work has extensively optimized EP communication. Libraries such as DeepEP~\cite{deepep2025}, HybridEP~\cite{hybridep2025}, and NCCL-EP~\cite{ncclep} have improved dispatch throughput and latency through fusing permutation with communication, and communication--computation overlap. While these techniques can reduce the number of coexisting intermediate buffers, token dispatch and expert computation still create memory peaks that limit training (\S~\ref{sec:motivation}). Multiple expanded activations can coexist as dispatch and compute stages allocate new outputs while retaining their inputs. Reducing this memory footprint requires rethinking how communication and computation are organized in the MoE block.

Tutel~\cite{tutel} pipelines token dispatch and expert computation using multiple streams, while FasterMoE~\cite{fastermoe2022} uses fine-grained scheduling to overlap communication with computation. Janus~\cite{janus2023} instead fetches expert weights to process local tokens. These works primarily target communication efficiency, while peak GPU memory during MoE execution remains a limiting factor, especially during backward.

In this paper, we present \name, a ring-based MoE execution model that reduces peak GPU memory by computing locally as data circulates through the ring. \name circulates expert weights or token activations through a ring of GPUs. At each hop, a GPU computes partial results using the data present on that device and forwards the visiting data to the next GPU; the partial results accumulate into the final output as the data completes the ring. Because no GPU ever needs to receive and hold data from all other GPUs, the memory footprint of the MoE block is capped by the local data plus a fixed number of in-flight partitions, independent of the top-$k$ factor and the EP degree.
Rather than reserving a fixed token capacity for each expert, \name processes the routed token--expert pairs partition by partition. This enables dropless, padding-free expert computation and supports longer sequences and larger batch sizes on the same hardware.

To turn these memory savings into higher training throughput, \name selects between expert and token routing according to communication volume and overlaps transfers with local computation. Memory efficiency must also extend to backward: activation checkpointing discards forward intermediates, but reconstructing them together can recreate a large memory peak. Reconstructed activations remain in memory while gradient computation produces additional intermediate tensors. The ring structure naturally supports memory-efficient MoE recomputation during backward: at each hop, \name reconstructs the intermediates needed for local gradient computation and releases them immediately afterward. This prevents expert intermediates from accumulating across hops and leaves more memory for retaining attention activations, reducing attention recomputation.

We make the following contributions:
\begin{itemize}[itemsep=2pt, topsep=2pt, parsep=0pt, partopsep=0pt]
\item We identify the memory overhead of full top-$k$-expanded dispatch buffers and propose a ring-based MoE execution model that avoids materializing them, with adaptive routing between experts and tokens.
\item We design memory-efficient MoE recomputation that reduces the backward memory peak without an additional ring traversal, and selective attention recomputation that converts these memory savings into higher training throughput.


\item We evaluate \name on production MoE models and across varied expert configurations. Compared with all-to-all, \name achieves a 2$\times$ average speedup in MoE layer execution at EP=8, reduces peak GPU memory during MoE execution by 7.6$\times$ on average at EP=64, and extends the largest tested trainable sequence length by up to 2.85$\times$ in full-model training.
\end{itemize}

\section{Background and Motivation }
\label{sec:motivation}
\subsection{MoE Training with Expert Parallelism}

\begin{figure}[!htbp]
    \centering
    \includegraphics[width=\linewidth]{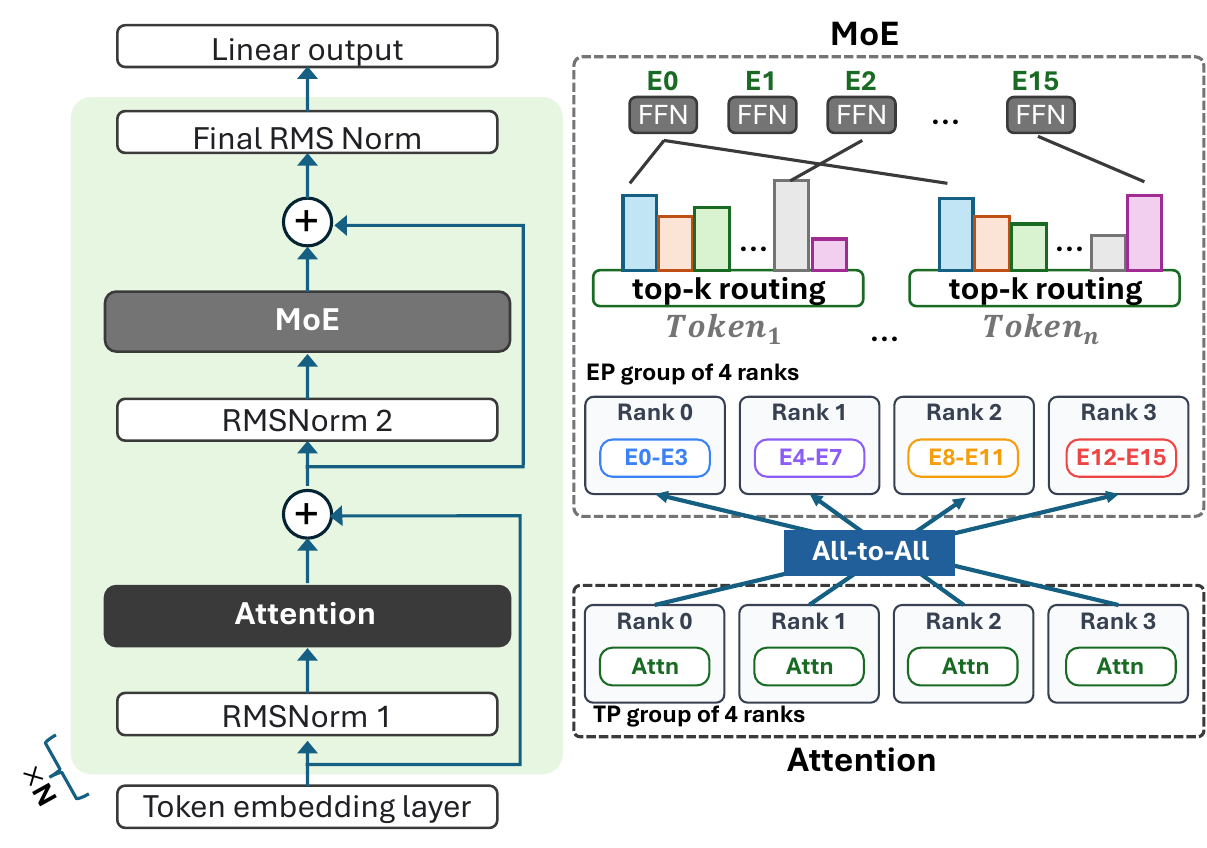}
    \caption{MoE-based transformer architecture with expert parallelism. Each layer consists of an attention block followed by a MoE block, where the router selects top-k experts per token. Experts are distributed across GPUs via expert parallelism, with tokens dispatched through all-to-all communication.}
    \vspace{-10pt}
    \label{fig:moe_arch}
\end{figure}

MoE-based language models replace the dense feedforward network (FFN) in each transformer layer with a set of expert FFNs and a learned router (Figure~\ref{fig:moe_arch}, left). For each input token, the router computes a score over all experts and selects the top-k highest-scoring experts to process that token (Figure~\ref{fig:moe_arch}, upper right). The outputs of the selected experts are weighted by the router scores and summed to produce the final output. Because only $k$ out of $E$ total experts are activated per token, MoE models scale total parameters without proportionally increasing per-token compute cost.

When the number of experts exceeds what a single GPU can hold, expert parallelism (EP) distributes experts across GPUs, with each GPU hosting a subset of experts (Figure~\ref{fig:moe_arch}, lower right). Because each token may be routed to experts residing on different GPUs, cross-GPU communication such as all-to-all is required to deliver tokens to their assigned experts and return results after computation.

\begin{figure*}[!htbp]
    \centering
    \includegraphics[width=\linewidth]{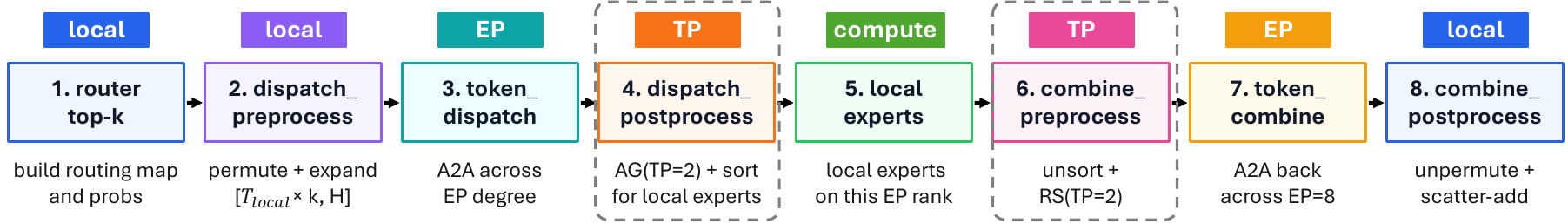}
    \caption{Stages of the all-to-all dispatcher in Megatron-LM. Each stage is annotated with its communication scope (local, EP, TP, or compute). The dispatch path expands tokens by top-k and redistributes them across EP ranks; the combine path reverses the process after expert computation. ``A2A'' and ``AG'' represents all-to-all and all-gather communication, respectively. Dashed stages are present only when tensor parallelism (TP$>$1) is enabled for expert computation.
    }
    \label{fig:dispatcher}
\end{figure*}

The standard mechanism for this communication is the \textit{all-to-all dispatcher}. Figure~\ref{fig:dispatcher} illustrates its stages. First, a dispatch preprocessing step permutes tokens by expert assignment and expands each token into k copies, producing a tensor of shape $[\text{tokens} \times k, H]$. This expanded tensor is then dispatched across GPUs via an all-to-all collective, so that each GPU receives the tokens assigned to its local experts. A postprocessing step reorganizes the received tokens for efficient expert computation via grouped GEMM. After expert computation, the reverse path combines results: a combine preprocessing step prepares outputs, a second all-to-all returns them to the originating GPUs, and a final unpermutation restores the original token order with weighted aggregation.


\subsection{Peak Memory Consumption in MoE Training}
\begin{figure}[!htbp]
    \centering
    \includegraphics[width=\linewidth]{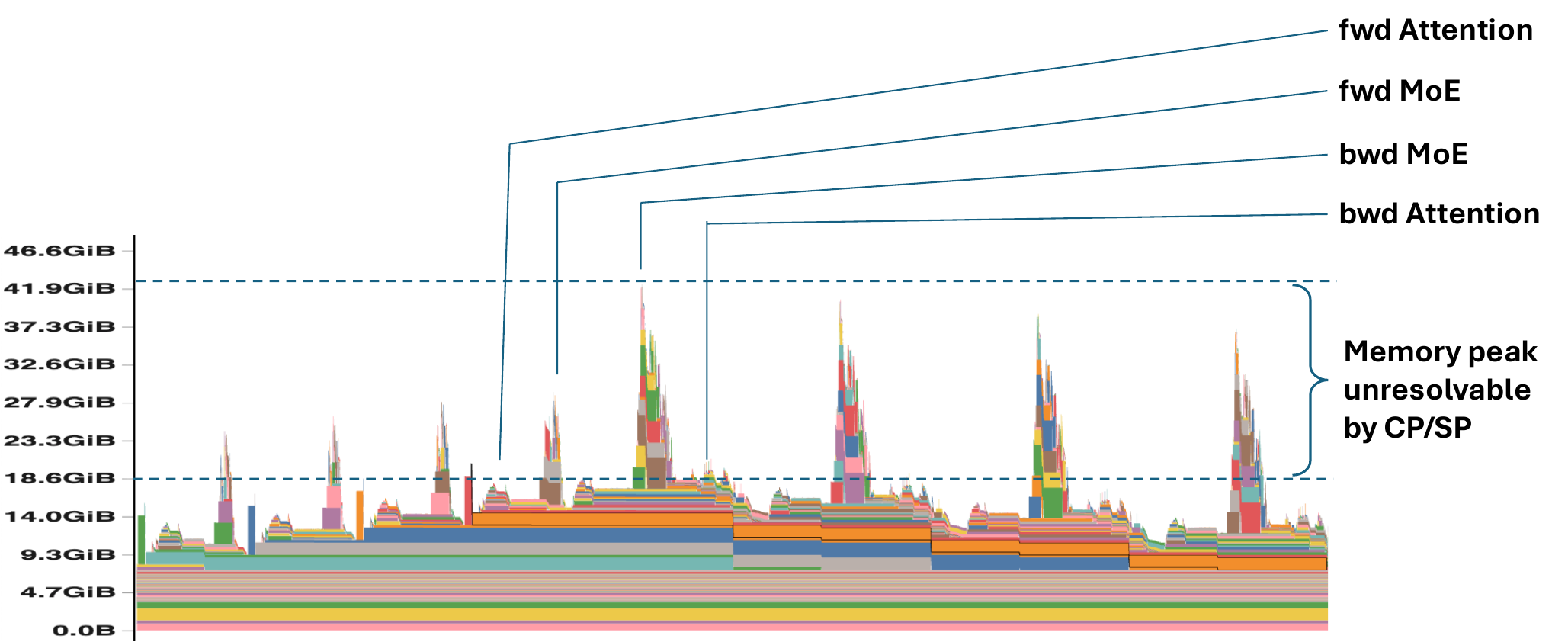}
    \caption{Memory profile during training of a Qwen3-style MoE model. MoE blocks create memory spikes in both forward and backward passes that dominate peak memory.}
    \vspace{-5pt}
    \label{fig:m1}
\end{figure}

\begin{figure}[!htbp]
    \centering
    \includegraphics[width=\linewidth]{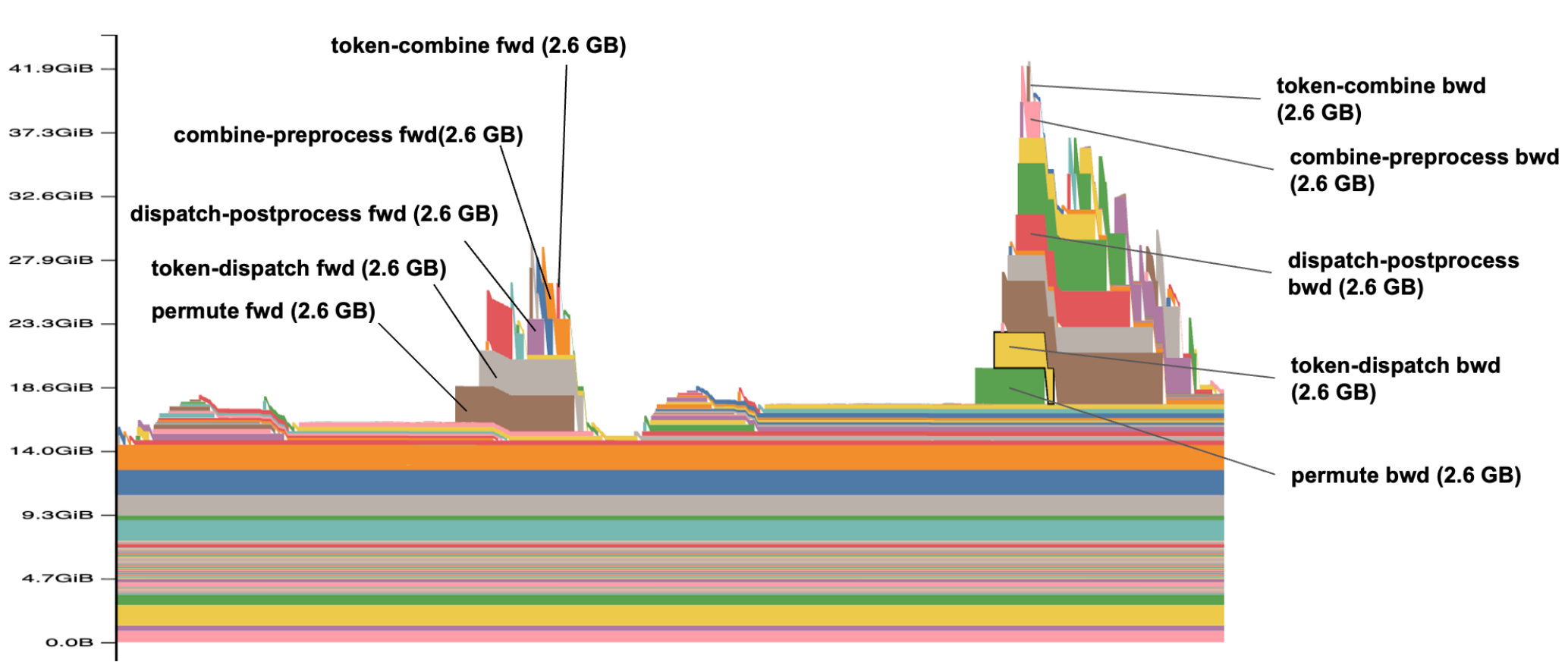}
    \caption{Memory profile of a single transformer layer. The MoE dispatch pipeline produces multiple 2.6 GiB intermediate activations (8$\times$ expansion of the base activation due to top-k routing). }
    \vspace{-5pt}
    \label{fig:m2}
\end{figure}

To understand where peak memory arises in distributed MoE training, we profile a Qwen3-style MoE model (Table~\ref{tab:models}) with results shown in Figure~\ref{fig:m1} and~\ref{fig:m2}. Each of the 4 transformer layers consists of a multi-head attention block followed by a MoE block with top-8 routing. We train with a micro-batch size of 24 and sequence length of 16384 on 8 H100 GPUs. For the attention portion of each layer, we apply tensor parallelism with sequence parallelism (TP=SP=2) and context parallelism (CP=4), which together reduce the local sequence length to 2048. For the MoE portion, we use expert parallelism (EP=8) with expert tensor parallelism disabled (ETP=1), following MoE Parallel Folding~\cite{moeparallelfolding2025} where attention and MoE adopt independent parallelism mappings. Activation recomputation is enabled throughout. This configuration represents a well-optimized baseline: existing attention memory techniques are fully applied, reducing the MoE input activation tensor on each GPU to 336 MiB\footnote{Per-GPU activation size = micro-batch $\times$ local sequence length $\times$ hidden dimension $\times$ 2 bytes (bf16) = $24 \times 2048 \times 3584 \times 2~\text{bytes} = 336~\text{MiB}$. }.

Figure~\ref{fig:m1} shows the memory profile across all 4 layers during training. Sharp spikes appear repeatedly in both forward and backward passes, reaching up to 51 GB while the memory between spikes remains around 14 GB. Each spike corresponds to a MoE block within a transformer layer. Although SP and CP reduce the number of tokens per GPU entering each MoE block to $B \times S / (\text{SP} \times \text{CP}) = 49{,}152$, the subsequent top-k expansion multiplies this by 8, producing the dispatch buffers that cause the spikes. As a result, it is the MoE blocks, not attention, that determine the out-of-memory (OOM) boundary for a given training configuration.

Figure~\ref{fig:m2} zooms into a single transformer layer during both forward and backward passes. The top-k-expanded dispatch tensors each occupy approximately 2.6 GiB, an 8$\times$ expansion from the 336 MiB input activation. The all-to-all dispatcher pipeline produces multiple such activations through token permutation, dispatch, and combine. These expanded tensors coexist as dispatch and combine stages allocate new outputs while retaining their inputs, driving the per-layer peak above 42 GB during backward.

\begin{figure*}[!htbp]
    \centering
     \includegraphics[width=\linewidth ]{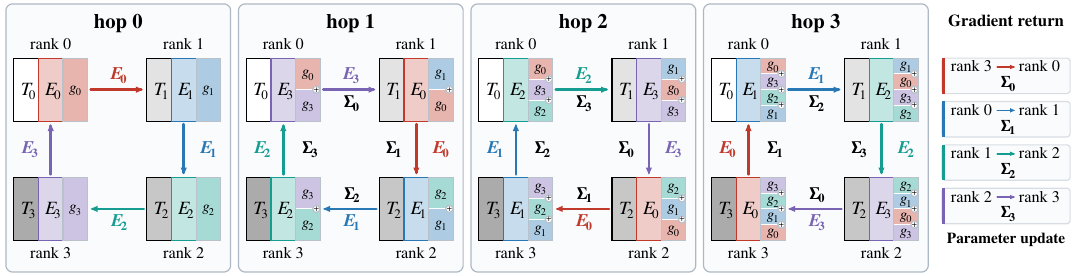}
    \vspace{-10pt}
\vspace{-5pt}
    \caption{Expert routing during backward at EP=4. Tokens $T_r$ stay on rank $r$. The colored stack shows contributions $g_j$ from visiting expert partition $E_j$ to the local token gradient. On the arrows, $\Sigma_j$ denotes the weight gradient of $E_j$, accumulated across ranks. Weight gradients follow their corresponding weights one hop later. The first transfer carries only weights; the final gradient-only transfer returns the completed weight gradients to their owners for parameter updates.}
    \label{fig:ring_moe}
\end{figure*}


These observations reveal that peak memory in distributed MoE training is driven by two factors. First, each intermediate activation in the dispatch pipeline is proportional to the number of tokens per GPU, the top-k routing factor, and the hidden dimension, all of which are growing in frontier architectures (Table~\ref{tab:trend}). Second, multiple such activations coexist in memory during execution of the dispatch pipeline. Together, these factors produce the memory spikes observed in Figure~\ref{fig:m1}. Addressing this bottleneck requires fundamentally redesigning how communication and computation are organized in the MoE block.


\section{Design}
\label{sec:design}

\name reduces peak GPU memory by computing locally as data circulates through a ring. We describe ring-based dispatch and its communication--computation overlap in Section~\ref{sec:local}, adaptive routing in Section~\ref{sec:adaptive}, memory-efficient MoE recomputation in Section~\ref{sec:saving_MoE_recompute}, and selective attention recomputation in Section~\ref{sec:adaptive_recompute}.

\subsection{Ring-Based MoE Dispatch}
\label{sec:local}

\name circulates data through the expert-parallel group in a ring instead of gathering all routed tokens in a single step (Figure~\ref{fig:ring_moe}). Both the experts and the tokens are partitioned across the EP group: each rank hosts a subset of the experts, as in standard expert parallelism, and holds the tokens of its local batch. We illustrate ring execution using expert routing; Section~\ref{sec:adaptive} discusses token routing and the choice between the two modes. At each hop, a GPU computes partial results using its local tokens and the \emph{visiting expert partition}, accumulates them into the output, and forwards the expert weights to its ring neighbor via peer-to-peer communication. After $N$ computation hops across the $N$ GPUs in the EP group, every partition has visited every GPU, so each token has been processed by all of its assigned experts and the output activation is complete.

\noindent\textbf{Forward execution.}
To illustrate, consider a forward hop using expert routing. The GPU uses the routing map to find which local tokens selected any of the visiting experts. For matching pairs, it locally permutes the relevant tokens, executes the expert GEMM, weights the outputs by the router probabilities, and adds the results to the running output activation. Tokens that did not select any visiting expert are skipped. Thus, each hop computes only the token--expert pairs selected by the router, while the expert partition follows the fixed ring schedule.

\noindent\textbf{Backward execution.}
With expert routing, each backward hop reconstructs the required expert intermediates and computes gradients (Section~\ref{sec:saving_MoE_recompute}). Gradients of the MoE outputs, token inputs, and router probabilities remain local. Expert-weight gradients are summed across ranks in reused communication buffers and returned to their owners before parameter updates.


Figure~\ref{fig:ring_moe} illustrates backward execution at EP=4. On rank 0, $T_0$ remains resident as $E_0$, $E_3$, $E_2$, and $E_1$ visit at hops 0--3. The colored stack represents in-place accumulation of the $g_j$ contributions into one token-gradient buffer; $\Sigma_j$ on the arrows denotes the accumulated weight gradient of $E_j$. At hop 0, computation with $E_0$ produces both $g_0$ for the local token gradient and the first contribution to $\Sigma_0$. Rank 0 sends only $E_0$ at hop 0, then $\Sigma_0$ together with $E_3$ at hop 1. Weights return to their owners at hop 3, followed by the final gradient-only return shown on the right before parameter updates.

\name is mathematically equivalent to all-to-all dispatch. Each token--expert pair computes the same GEMM with the same weights and the same router probabilities. The difference is the execution order: all-to-all gathers the pairs at the destination GPU before computing, while \name accumulates partial outputs incrementally across hops.

\label{sec:overlap}
\noindent\textbf{Communication--computation overlap.}
Ring-based dispatch replaces the all-to-all dispatch pipeline with a sequence of peer-to-peer hops. Although this avoids the full expanded dispatch buffers, performing computation and communication sequentially at every hop would add substantial latency. Overlapping them is therefore critical to making ring-based dispatch practical. Forward requires overlap between local computation and weight transfer; backward additionally requires weight gradient transfer. We describe the schedule with expert routing below. Token routing follows the same principle, overlapping transfer of the next token partition with computation on the current partition.

In the forward pass, while a GPU computes using the currently visiting expert partition and its local tokens, it prefetches the next partition's weights from its neighbor via asynchronous P2P communication. If this transfer completes before the current computation finishes, the next partition is available without a communication stall. With effective overlap, the per-hop time approaches the larger of computation and transfer time, rather than their sum.

During backward, \name overlaps local recomputation and gradient computation with merged transfers of expert weights and accumulated weight gradients. Gradients follow their corresponding weights one hop later, so each transfer combines weights and gradients from different expert partitions.

The key observation enabling this overlap is that local partial weight gradients can be computed without waiting for the incoming accumulated gradient buffer. After local gradient computation completes, the GPU synchronizes with the incoming buffer, adds its local partial gradient in-place, and asynchronously forwards the updated buffer to the next GPU. If the incoming buffer has not arrived, the GPU waits at this synchronization point. With effective overlap, the per-hop backward time approaches the longer of local computation and communication, plus synchronization and in-place accumulation. Both forward and backward expert computation can adopt optimized GEMM kernels such as GroupedGEMM. Section~\ref{sec:eval_analysis} evaluates the contributions of overlap and GroupedGEMM separately.

\noindent\textbf{Memory consumption.}
Each GPU retains its resident data, output or gradient accumulators, and the current and prefetched partitions with their communication buffers. Local permutation and expert intermediates cover only the token--expert pairs matched at the current hop; their size still depends on top-$k$ and routing. \name avoids gathering the full top-$k$-expanded activations across partitions or constructing a full expanded buffer during combine. This removes the coexisting dispatch buffers identified in Section~\ref{sec:motivation}, while preserving the same token--expert computation. Section~\ref{sec:saving_MoE_recompute} explains how reconstructing and consuming expert intermediates within each hop also reduces their lifetime and peak memory during backward.

\begin{figure}[!htbp]
    \centering
    \includegraphics[width=0.9\linewidth]{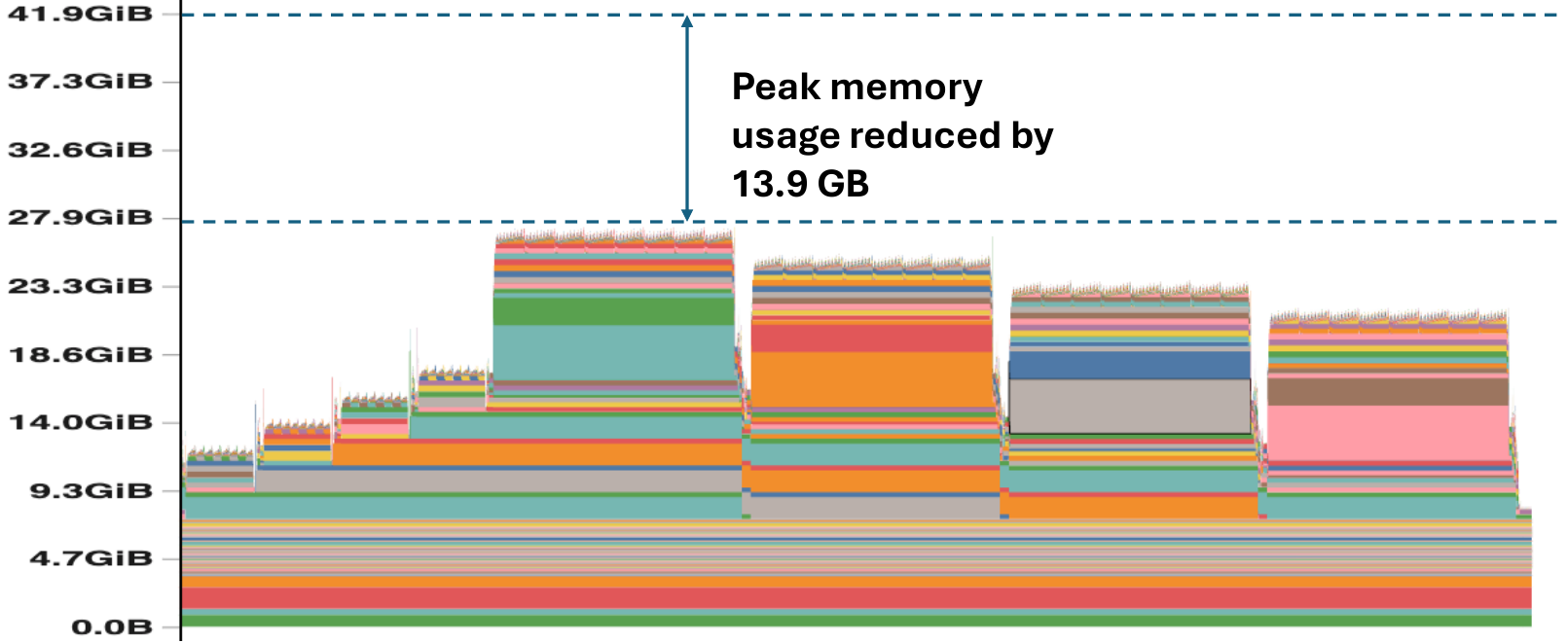}
        \vspace{-10pt}
    \caption{Memory profile under \name with the same training configuration as Figure~\ref{fig:m1}. The dashed lines mark the all-to-all and \name peak levels, with a reduction of 13.9 GiB. The sharp MoE memory spikes are substantially reduced.}
    \vspace{-10pt}
    \label{fig:m3}
\end{figure}

\begin{figure}[!htbp]
    \centering
    \includegraphics[width=0.9\linewidth]{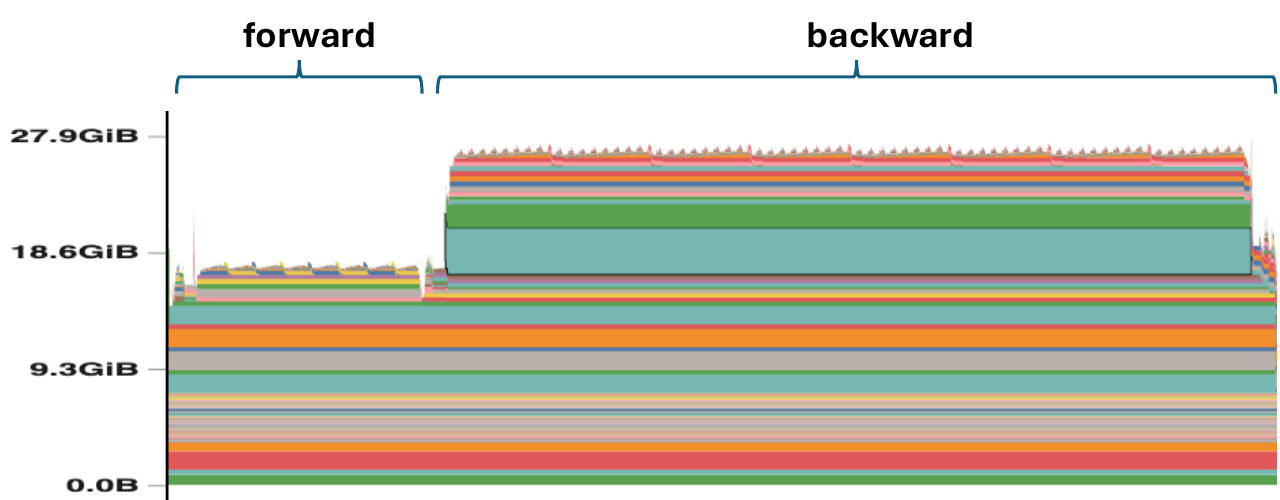}
    \caption{Forward and backward GPU memory profile of one transformer layer under \name, corresponding to Figure~\ref{fig:m2}.}
    \vspace{-5pt}
    \label{fig:m4}
\end{figure}

Figure~\ref{fig:m3} shows the resulting memory profile under the same training configuration as Figure~\ref{fig:m1}. The sharp MoE memory spikes are substantially reduced, and peak memory is reduced by 13.9 GiB. Figure~\ref{fig:m4} zooms into one layer's forward and backward execution, corresponding to Figure~\ref{fig:m2}. The coexisting 2.6 GiB dispatch buffers no longer appear; instead, local expert computation produces smaller memory fluctuations across hops. Backward still uses more memory than forward, but it no longer exhibits the large dispatch-buffer spike seen with all-to-all.

This design eliminates the need for token dropping and capacity-factor padding to bound dispatch buffers. In all-to-all dispatch, routing imbalance can concentrate tokens on the GPU that hosts a popular expert. With expert routing, each GPU instead processes its fixed local tokens as expert partitions visit, regardless of which experts the tokens select. This balances resident token memory across GPUs, although uneven token--expert matches can still leave ranks idle within a hop. For architectures with shared experts such as Qwen2-MoE, shared experts are computed separately outside the ring, as they are replicated on every GPU and do not require dispatch communication.

\subsection{Adaptive Routing: Experts vs Tokens}
\label{sec:adaptive}

\name can circulate either expert weights or token activations through the ring while the other remains resident on each GPU. Both modes produce identical results (Section~\ref{sec:local}), but differ in communication volume (Table~\ref{tab:routing}). With token routing, each token partition carries its routing metadata and accumulated outputs back to its original GPU. During backward, the token inputs, MoE output gradients, and accumulated token-input and router-probability gradients circulate together, while expert-weight gradients accumulate locally. Both modes reuse communication buffers across hops.

Considering expert parallelism alone, the comparison involves two factors: per-hop communication volume and the tokens or expert weights kept resident on each GPU (both modes traverse the same $N$ GPUs). We use the notation in Table~\ref{tab:routing}: $H$ is the hidden dimension, $W$ is the size of one expert's weights, and $M$ is the routing metadata size per token. In forward, expert routing communicates $E_{\text{local}} \times W$ per hop while keeping the local tokens resident. Token routing communicates the local token partition and its routing metadata while keeping $E_{\text{local}}$ experts resident. Expert routing is preferred when token activations are large relative to expert weights: the communication volume is smaller and the tokens, which would be expensive to move, remain stationary. Token routing is preferred when expert weights are large relative to token activations: the communication volume is smaller and the large expert weights remain stationary.

Table~\ref{tab:routing} includes gradient communication during backward. For expert routing, it reports the merged payload of expert weights and accumulated weight gradients described in Section~\ref{sec:local}.

\begin{table}[!htbp]
\centering
\caption{Per-hop communication volume and resident memory for expert routing vs token routing. $E_{\text{local}}$: experts per GPU, $W$: single expert weight size, $B$: micro-batch, $S$: sequence length, $H$: hidden dimension, $M$: per-token routing metadata size.}
\label{tab:routing}
\small
\begin{tabular}{lcc}
\toprule
 & \textbf{Routing Expert} & \textbf{Routing Token} \\
\midrule
\rowcolor{gray!15}
\multicolumn{3}{l}{\textit{Basic (EP only, N = EP hops)}} \\[2pt]
Resident       & Tokens: $B \times S \times H$       & Experts: $E_{\text{local}} \times W$ \\[4pt]
Comm. (fwd)    & $E_{\text{local}} \times W$          & $B \times S \times (H + M)$ \\[4pt]
Comm. (bwd)    & $2 \times E_{\text{local}} \times W$ & $2 \times B \times S \times (H + M)$ \\
\midrule
\rowcolor{gray!15}
\multicolumn{3}{l}{\textit{With SP, CP, and ETP}} \\[2pt]
Resident       & $\frac{B \times S}{\text{SP} \times \text{CP}} \times H$                  & $\frac{E}{\text{EP}} \times \frac{W}{\text{ETP}}$ \\[6pt]
Comm. (fwd)    & $\frac{E}{\text{EP}} \times \frac{W}{\text{ETP}}$                         & $\frac{B \times S}{\text{SP} \times \text{CP}} \times (H + M)$ \\[6pt]
Comm. (bwd)    & $2 \times \frac{E}{\text{EP}} \times \frac{W}{\text{ETP}}$                & $2 \times \frac{B \times S}{\text{SP} \times \text{CP}} \times (H + M)$ \\
\bottomrule
\end{tabular}
\end{table}

In distributed MoE training, EP typically works alongside other parallelism dimensions that further adjust both resident memory and communication volume. Sequence parallelism (SP) and context parallelism (CP) partition the sequence dimension, reducing tokens per GPU to $B \times S / (\text{SP} \times \text{CP})$. Expert tensor parallelism (ETP) splits each expert's weight matrices across GPUs, reducing per-expert weight size to $W / \text{ETP}$. MoE Parallel Folding~\cite{moeparallelfolding2025} allows ETP to be configured independently of attention-layer TP. Both adjustments apply to both modes, as shown in the second section of Table~\ref{tab:routing}. Because these adjustments can shift the balance in either direction, the preferred routing mode depends on the full training configuration rather than the model architecture alone. For example, a model with many experts may favor expert routing under high ETP (which shrinks expert volume) but token routing under low ETP. \name evaluates both volumes at training start using the full configuration and selects the mode with lower per-hop communication.

As shown in Table~\ref{tab:trend}, frontier MoE models have grown from 8 experts (Mixtral) to several hundred (Qwen3.5, Kimi K2). At a fixed EP degree, this increases the number of experts per GPU $E/EP$. For a fixed expert size, more experts increase the weight volume transferred at each hop under expert routing. Token routing avoids transferring these expert weights, making it an attractive option when expert partitions are large relative to token activations. The choice still depends on expert width and token count.

The physical topology of the ring also matters. When EP spans multiple nodes, cross-node hops use inter-node links (e.g., RDMA) with substantially lower bandwidth than intra-node NVLink. All-to-all dispatchers address this through hierarchical algorithms and optimized kernels such as Tutel and DeepEP that pipeline cross-node transfers. \name's ring structure instead serializes communication into P2P hops, which creates opportunities to overlap each hop's transfer with local computation (Section~\ref{sec:local}), but requires careful scheduling to avoid underutilizing bandwidth on the fast intra-node hops. Its memory benefit comes from avoiding the full expanded dispatch buffers, while its latency depends on the balance between local computation and transfers over the available links. 

\subsection{Memory-Efficient MoE Recomputation}
\label{sec:saving_MoE_recompute}

\begin{figure}[!htbp]
    \centering
    \includegraphics[width=\linewidth]{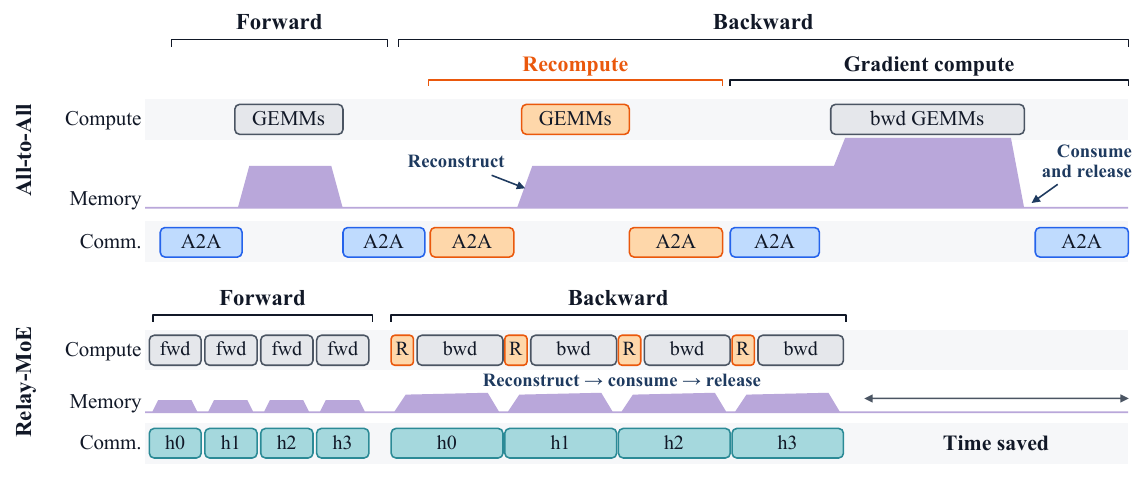}
    \vspace{-10pt}
    \caption{MoE recomputation with all-to-all and \name (EP=4). R denotes recomputation; h0--h3 denote ring hops. Purple profiles show only intermediate activations and temporary gradients. The final gradient-only transfer is not shown.}
\vspace{-10pt}
    \label{fig:recompute_MoE_saving}
\end{figure}

Figure~\ref{fig:recompute_MoE_saving} compares MoE recomputation with all-to-all and \name. In the upper timeline, activation checkpointing~\cite{chen2016} retains the inputs needed to replay the MoE layer during backward. Reconstructed activations remain live while gradient computation produces additional temporary tensors. The memory profile therefore rises during recomputation and again during gradient computation, producing a backward peak higher than the forward peak.


The ring structure naturally supports recomputation within each backward hop, as shown in the lower timeline. With expert routing, \name uses the saved local tokens and routing information together with the visiting expert partition to reconstruct the required intermediates (\texttt{R}). It immediately computes and accumulates the corresponding gradients, then releases the reconstructed intermediates before the next hop. Gradient computation still raises memory within a hop, but expert intermediates do not accumulate across hops, so their peak memory is determined by the largest hop. Saved inputs and gradient accumulators remain live across hops; weight and gradient communication buffers are reused. This schedule reduces the memory needed for expert computation during backward itself, beyond the savings from discarding forward activations.


Recomputation also reuses the data movement already required by backward. The visiting weights and local tokens are available for both reconstruction and gradient computation, so reconstructing expert intermediates requires no additional ring traversal. Token routing follows the same principle with visiting tokens and resident experts. The local recomputation participates in the overlap described in Section~\ref{sec:local}; it still adds computation, and its runtime cost depends on the balance between computation and communication.

\subsection{Selective Attention Recomputation}
\label{sec:adaptive_recompute}

The memory saved by \name can also reduce recomputation outside the MoE layer. We retain the MoE recomputation schedule above and use the available memory to selectively retain attention activations. These activations can then be reused during backward, reducing attention recomputation and improving training throughput.

For each attention layer $i$, recomputation option $j$ has a memory cost $m_{i,j}$ for retained activations and a compute saving $c_{i,j}$ from avoiding the corresponding backward recomputation. Both are estimated from the model configuration and parallelism settings. For grouped-query attention with FlashAttention, the activation sizes depend on the micro-batch size, local sequence length, number of attention heads, and head dimension. The compute saving also depends on sequence length and the attention mechanism used by each block.

\name uses the difference between GPU capacity and peak memory under full recomputation as its memory budget. It ranks the available options by their compute saving per unit of memory $c_{i,j}/m_{i,j}$ and greedily selects options within this budget. The selection is computed once at the start of training and applied throughout. 

\section{Evaluation}
\label{sec:evaluation}

\subsection{Experimental Setup}
\label{sec:setup}

\noindent\textbf{Platform.}
All experiments were run on MareNostrum~5 in the Accelerated (ACC) partition.
Each node contains four NVIDIA H100 GPUs (64\,GB HBM2e per GPU) and two Intel Xeon Platinum 8460Y+ CPUs (40 cores each).
GPUs within a node are interconnected via NVLink~4.0, and nodes are connected through an NDR InfiniBand network with a fat-tree topology~\cite{MN5}.

\noindent\textbf{Software and baselines.}
We implement \name on top of Megatron-LM~\cite{megatron-lm} in 2.3K lines of code.
For the MoE layer experiments, we compare \name against FasterMoE~\cite{fastermoe2022}, Tutel~\cite{tutel}, and Megatron-LM's all-to-all dispatcher, varying token counts and expert parallelism degrees.
The All-to-All baseline uses the default MoE dispatcher from a recent version of Megatron-LM (May 2026).
For end-to-end training, we use the same All-to-All implementation as our primary baseline.
We also enable DeepEP~\cite{deepep2025} within Megatron-LM, which optimizes token dispatch and combine while retaining the same all-to-all formulation.

\noindent\textbf{Models.}
We evaluate \name using Qwen3-style configurations that vary the number of experts (64--384), the routing degree (top-4 to top-16), and the expert FFN width (512--2048). We adjust the number of layers to keep the total trainable parameter count comparable across settings, as shown in Table~\ref{tab:models}. Single-layer experiments isolate MoE execution, while full-model results also reflect the different layer counts. Unless stated otherwise, these experiments use 2 nodes (8 GPUs) with TP=4, CP=2, PP=1, and EP=8.

We also evaluate production-model training with Qwen2-57B-A14B, Qwen3-30B-A3B, and Qwen3.5-35B-A3B on 8 nodes (32 GPUs). We enable NVIDIA's MoE Parallel Folding~\cite{moeparallelfolding2025}, a state-of-the-art parallelization strategy that allows attention and MoE layers to use independent configurations. We use TP=2, CP=4, PP=4, and EP=8 to evaluate the benefits of \name on top of these parallelism optimizations.

\begin{table}[t]
\centering
\caption{Model configurations used in evaluation. $L$: layers, $E$: experts, $K$: top-k. Scaling study configs use Qwen3-style architecture (SwiGLU, RMSNorm, RoPE) with PP=1. $L$ varies to keep the number of trainable parameters similar per GPU. }
\vspace{-5pt}
\label{tab:models}
\small
\begin{tabular}{lcccccc}
\toprule
\textbf{Name} & \textbf{L} & \textbf{Hidden} & \textbf{FFN\textsubscript{moe}} & \textbf{E} & \textbf{K} & \textbf{Params} \\

\midrule
\rowcolor{gray!15}
\multicolumn{7}{l}{\textit{Production architectures (end-to-end evaluation)}} \\
\scriptsize{Qwen3-30B-A3B}  & 48 & 2048 & 768  & 128   & 8  & 30B \\
\scriptsize{Qwen3.5-35B-A3B}  & 40 & 2048 & 512  & 256  & 8  & 35B \\
\scriptsize{Qwen2-57B-A14B}  & 28 & 3584 & 2560  & 64  & 8  & 57B \\
\midrule
\rowcolor{gray!15}
\multicolumn{7}{l}{\textit{Varying expert count}} \\
MoE-E64    & 24 & 2048 & 768  & 64  & 8  & 7.8B \\
MoE-E128   & 12 & 2048 & 768  & 128 & 8  & 7.6B \\
MoE-E384   & 4  & 2048 & 768  & 384 & 8  & 7.5B \\
\midrule
\rowcolor{gray!15}
\multicolumn{7}{l}{\textit{Varying top-k}} \\
MoE-K4     & 12 & 2048 & 768  & 128 & 4  & 7.5B \\
MoE-K8     & 12 & 2048 & 768  & 128 & 8  & 7.6B \\
MoE-K12    & 12 & 2048 & 768  & 128 & 12 & 7.6B \\
MoE-K16    & 12 & 2048 & 768  & 128 & 16 & 7.6B \\
\midrule
\rowcolor{gray!15}
\multicolumn{7}{l}{\textit{Varying expert FFN width}} \\
MoE-H512   & 18 & 2048 & 512  & 128 & 8  & 7.72B \\
MoE-H768   & 12 & 2048 & 768  & 128 & 8  & 7.6B \\
MoE-H2048  & 5  & 2048 & 2048 & 128 & 8  & 8.2B \\

\bottomrule
\end{tabular}
\vspace{-10pt}
\end{table}

\begin{figure*}[!htbp]
    \centering
    \includegraphics[width=\linewidth]{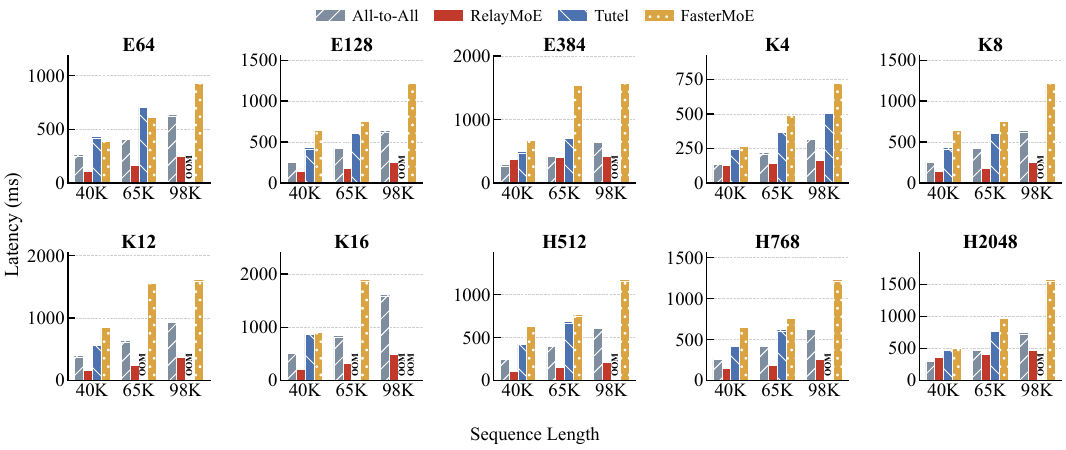}
    \vspace{-20pt}
    \caption{Latency of a single MoE layer at different sequence lengths, with EP=8 and micro-batch size 12.}
    \vspace{-5pt}
    \label{fig:baseline_seq_scaling}
\end{figure*}

\begin{figure*}[!htb]
    \centering
    \includegraphics[width=\linewidth]{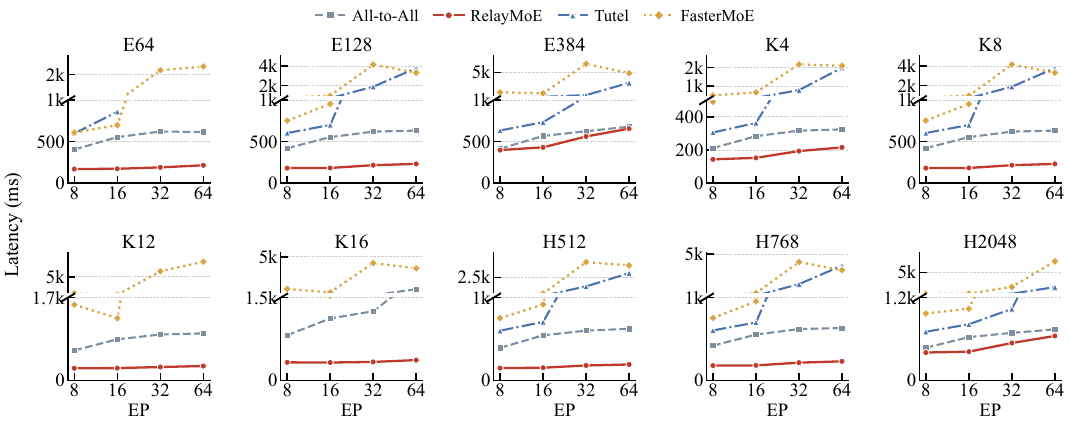}
    \vspace{-20pt}
    \caption{Forward-and-backward latency of a single MoE layer as EP increases from 8 to 64, with micro-batch size 12 and sequence length 8192 fixed per GPU. Missing Tutel points indicate OOM.}
    
    \label{fig:baseline_ep_scaling}
\end{figure*}
\noindent\textbf{Metrics.}
The MoE layer experiments report forward-and-backward latency, including communication. Average speedups are arithmetic means over distinct test configurations. For full-model experiments, we report model throughput in TFLOP/s/GPU, excluding recomputation FLOPs but including recomputation time. For a given model and sequence length, higher throughput means more training tokens processed per second.

\subsection{MoE Layer Performance}
\label{sec:eval_moe}

To understand the performance impact of MoE execution, we compare the forward-and-backward latency of \name with baselines. All comparisons use the same model, parallelism, batch size, sequence length, and recomputation settings.

\noindent\textbf{Latency comparison.} We study MoE execution latency using a single SwiGLU MoE layer from each configuration in Table~\ref{tab:models}. 
Figure~\ref{fig:baseline_seq_scaling} shows results at sequence lengths of 40K, 65K, and 98K. All systems use EP=8 on 8 GPUs, batch size $b=12$, and the same recomputation configuration.

\name outperforms all-to-all by 2$\times$ on average. For \texttt{E128}, which matches the expert configuration of Qwen3-30B-A3B, the speedup increases from 1.77$\times$ at 40K to 2.4$\times$ at 98K. Expert routing keeps tokens local and reduces communication volume at long sequence lengths. Longer sequences and larger micro-batches also increase the computation within each hop, helping amortize transfers and kernel launches.

For \texttt{E384} and \texttt{H2048}, \name is slower than all-to-all at 40K but becomes faster at 65K, reaching speedups of 1.51$\times$ and 1.56$\times$, respectively, at 98K. Their larger expert counts or wider FFNs increase the cost of transferring expert weights. As more tokens are processed, this weight-transfer volume stays fixed while all-to-all communication grows. The resulting crossover shows that the benefit of expert routing depends on the amount of token computation available to amortize weight transfers. Section~\ref{sec:eval_analysis} evaluates token routing for configurations with larger expert-weight volumes.

\begin{figure*}[!htbp]
    \centering
    \includegraphics[width=0.9\linewidth]{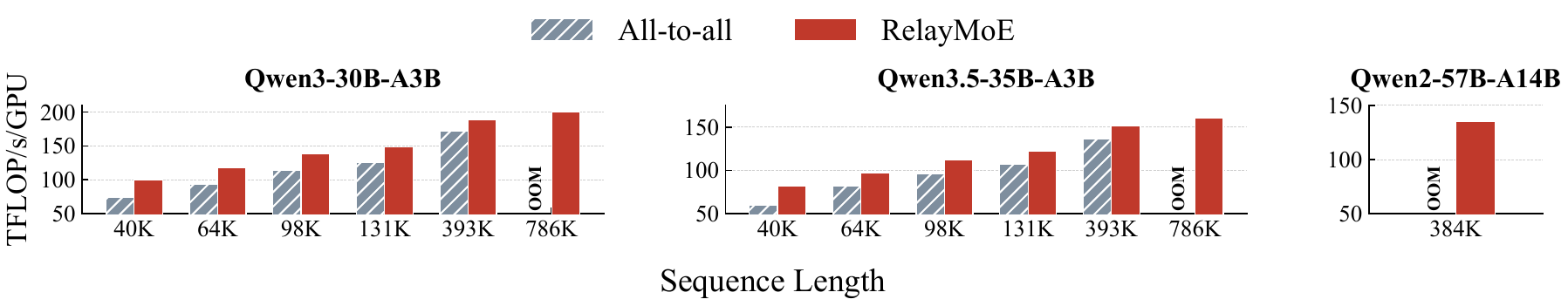}
    \vspace{-10pt}
    \caption{End-to-end training throughput on Qwen3-30B-A3B, Qwen3.5-35B-A3B, and Qwen2-57B-A14B. Each dispatcher uses its largest feasible micro-batch.}
    \vspace{-10pt}
    \label{fig:overall_perf}
\end{figure*}
\begin{figure*}[!htbp]
    \centering
    \includegraphics[width=\linewidth]{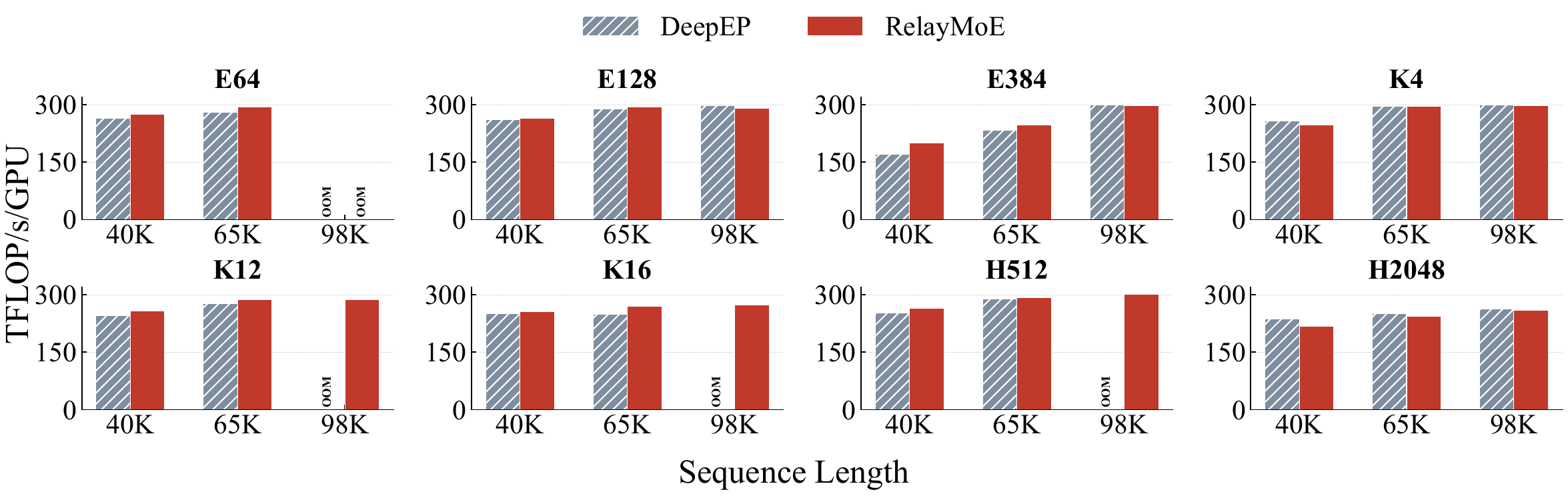}
    \vspace{-15pt}
    \caption{End-to-end training throughput with \name and DeepEP at EP=4. Where DeepEP fits, both use the same number of recomputed attention blocks, while \name uses its largest feasible micro-batch.}
    \vspace{-5pt}
    \label{fig:deepep}
\end{figure*}
Compared to FasterMoE, \name is 4.03$\times$ faster on average, with speedups ranging from 1.41 to 6.20$\times$. FasterMoE uses fine-grained scheduling to overlap token exchange with expert computation. In \name, expert routing keeps the transfer volume fixed as sequence length increases, while more tokens provide more local computation to overlap with these transfers. This helps amortize communication over longer sequences.

Compared to Tutel, \name is 2.94$\times$ faster on average. Tutel pipelines token dispatch and expert computation, but the communication volume of its expert-parallel execution still grows with sequence length and top-$k$. Tutel encounters OOM in several configurations at 65K and 98K, where more routed token assignments require larger dispatch buffers and more memory for expert intermediate activations.

\noindent\textbf{EP scaling performance.} We study whether \name and the baselines maintain their execution efficiency when processing proportionally more tokens at higher EP degrees. Figure~\ref{fig:baseline_ep_scaling} compares latency as EP increases from 8 to 64, with $b=12$ and sequence length 8192 fixed per GPU.

\name outperforms all-to-all at all tested EP degrees, with its speedup generally increasing as EP grows. For \texttt{K16}, the speedup increases from approximately 2.5$\times$ at EP=8 to 5$\times$ at EP=64. Expert routing reduces communication volume in these configurations and overlaps expert transfers with local computation.

In these single-layer experiments, \name reduces peak GPU memory during MoE execution by an average of 3.8$\times$ at EP=8 and 7.6$\times$ at EP=64, reaching 13.4$\times$ for \texttt{K16}. Increasing EP reduces the expert partition and the intermediate activations processed at each hop. These savings target the MoE execution peak identified in Figure~\ref{fig:m1}; Section~\ref{sec:eval_memory} examines their effect on full-model training, which also retains model states and activations from other layers.

Compared to FasterMoE and Tutel, \name's latency advantage generally increases with EP, reaching speedups of up to 27$\times$ and 16$\times$, respectively, at EP=64. \name reduces communication volume and overlaps expert transfers with local computation, helping limit latency growth at larger EP degrees. Tutel encounters OOM in some configurations due to buffer memory overhead. FasterMoE and Tutel are also slower than Megatron-LM all-to-all in all tested cases.

\subsection{Model Training Performance}
\label{sec:eval_end_to_end}

MoE-layer latency captures only part of \name's effect on training: lower MoE memory usage also leaves more space for attention activations and input tokens. We further quantify end-to-end throughput and trainable sequence lengths. 

\begin{figure*}[!htbp]
    \centering
    \includegraphics[width=\linewidth]{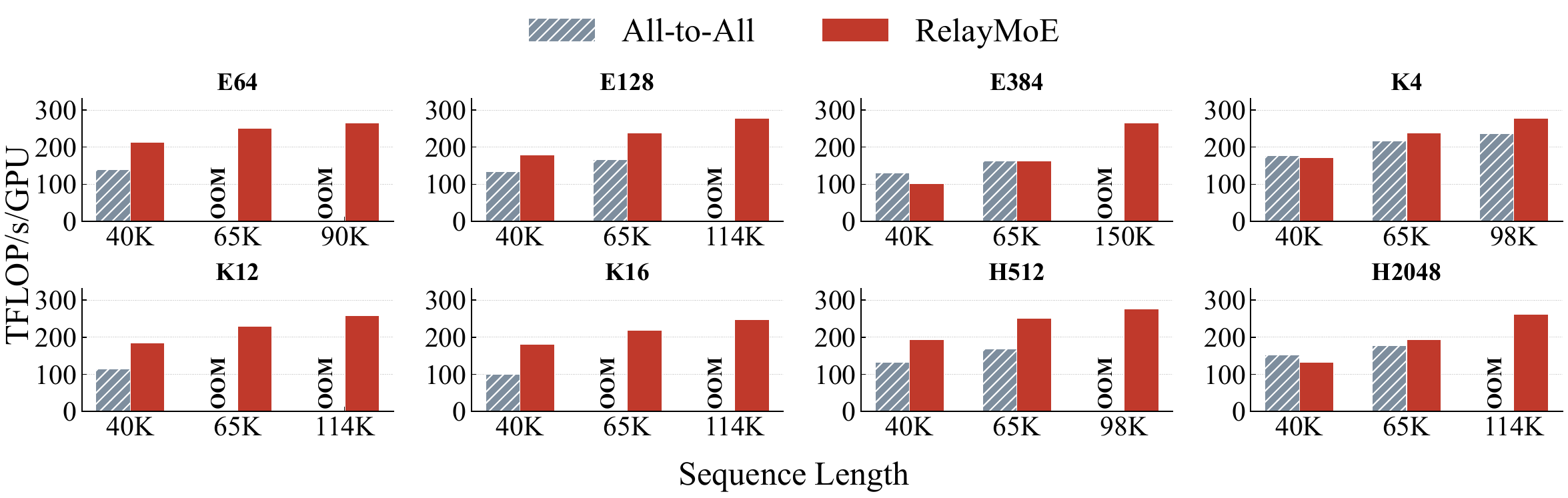}
    \vspace{-15pt}
    \caption{Training throughput with reduced attention recomputation. Sequence lengths 40K, 65K, and 98K use micro-batch sizes 12, 8, and 5, respectively.}
    \vspace{-5pt}
    \label{fig:recompute_scal}
\end{figure*}

\noindent\textbf{End-to-end performance on production models.}
We evaluate throughput and trainable sequence length on production MoE architectures. We compare \name with all-to-all on Qwen3-30B-A3B, Qwen3.5-35B-A3B, and Qwen2-57B-A14B using 32 GPUs with TP=2, CP=4, PP=4, and EP=8. At each sequence length, we select the largest micro-batch size for each dispatcher that fits in GPU memory. Figure~\ref{fig:overall_perf} shows that \name achieves higher throughput on both Qwen3-30B and Qwen3.5-35B at every sequence length where both dispatchers run, with the memory savings allowing larger micro-batches. The largest tested trainable sequence length increases from 393K to 786K, a 2$\times$ improvement. These gains hold across different expert granularities: Qwen3.5-35B doubles the expert count with a narrower expert FFN. On Qwen2-57B, \name trains at 384K while all-to-all runs out of memory, demonstrating the capacity benefit on this wider-expert model. Attention and the remaining training work limit how much faster MoE execution improves end-to-end throughput; the larger feasible micro-batches and contexts provide an additional benefit from lower memory usage.

\noindent\textbf{Comparison with DeepEP.}
We next compare \name with DeepEP to evaluate training throughput and capacity with optimized all-to-all communication. We use the Qwen3-style full models in Table~\ref{tab:models}, including their different layer counts, and evaluate sequence lengths 40K, 65K, and 98K. Both dispatchers use EP=4 within one node; inter-node DeepEP was unavailable with the platform's NVSHMEM and NIC configuration. Where DeepEP fits, it uses a fixed batch size and the smallest number of recomputed attention blocks that fits in GPU memory. \name uses the same number of recomputed attention blocks and its largest feasible micro-batch. Figure~\ref{fig:deepep} reports the results.

When both dispatchers fit in memory, \name generally achieves comparable training throughput while supporting larger micro-batches. Matching the number of recomputed attention blocks excludes the reduced-recomputation benefit. The larger micro-batch does not guarantee higher throughput: \texttt{H2048} remains slower for \name, although the gap narrows at longer sequences, consistent with the higher expert-transfer cost of its wide FFNs.

DeepEP runs out of memory on \texttt{K8}, \texttt{K12}, \texttt{K16}, and \texttt{H512} at 98K, while \name trains these configurations. DeepEP optimizes token dispatch and combine while retaining the all-to-all formulation. Optimized dispatch kernels retain the routed-token activation buffers, so faster communication alone does not remove their memory cost~\cite{memfine2025}. Here, lower memory usage expands the trainable configurations even when throughput gains are limited. At 98K, both dispatchers still run out of memory on \texttt{E64}, whose 24-layer configuration retains attention and checkpoint state across more layers; fewer experts per layer therefore do not guarantee a smaller full-model memory footprint.

\subsection{Impact of Memory Savings}
\label{sec:eval_memory}

The end-to-end comparisons combine changes in MoE execution with the effects of lower memory usage. We examine three uses of the saved memory: retaining attention activations to reduce recomputation, increasing the micro-batch, and extending the sequence length. The sequence-length comparison holds the micro-batch and full recomputation settings fixed, showing whether the capacity benefit persists without either of the first two adjustments.

\noindent\textbf{Recompute reduction for improved throughput.}

We first evaluate whether the memory savings of \name can be translated into higher training throughput by reducing attention activation recomputation. We use the Qwen3-style configurations from Table~\ref{tab:models} on 8 GPUs and evaluate three long-context settings with sequence lengths 40K, 65K, and 98K. For these three sequence lengths, we use micro-batch sizes 12, 8, and 5, respectively, which keeps the number of tokens per step approximately constant. For each model configuration and each dispatcher, we calculate the number of recomputed attention blocks from the available GPU memory budget using the method in Section~\ref{sec:adaptive_recompute}.

Figure~\ref{fig:recompute_scal} shows higher throughput for \name in most configurations, with speedups of up to 2.02$\times$ on \texttt{K16} at 40K. The saved MoE memory allows more attention activations to be retained for backward, reducing attention recomputation and improving throughput. The near-parity on \texttt{K4} is consistent with its lower top-$k$ expansion, which means less memory can be saved by removing dispatch buffers. 

\begin{figure}[!htbp]
    \centering
    \includegraphics[width=\linewidth]{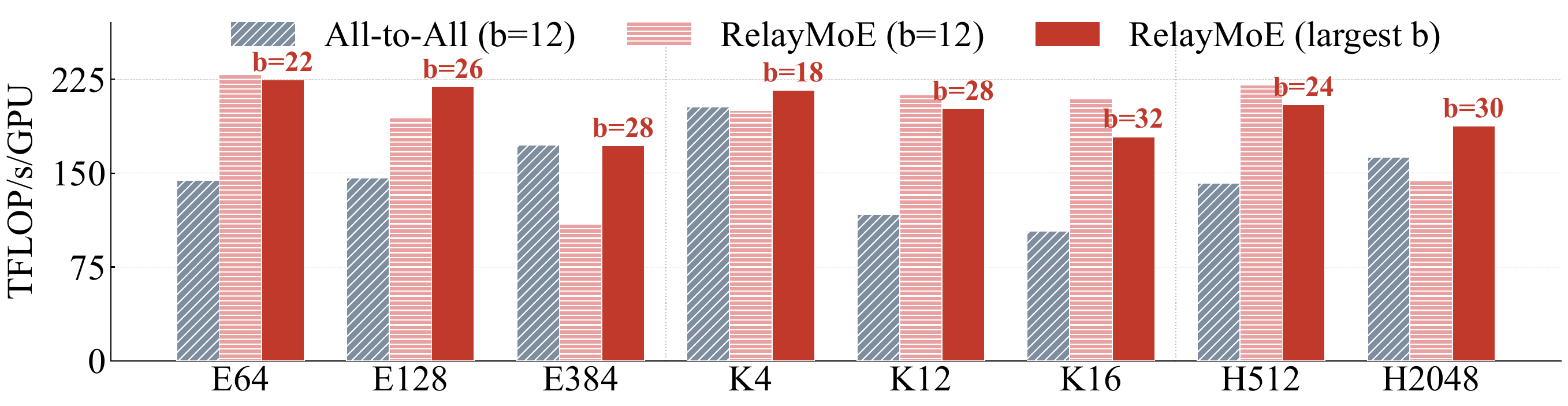}
    \vspace{-10pt}
    \caption{Training throughput at sequence length 40{,}960 with different micro-batch sizes. Labels above the dark-red bars give \name's largest feasible micro-batch sizes.}
    \vspace{-5pt}
    \label{fig:batch_scaling}
\end{figure}

All-to-all remains faster on \texttt{E384} at all three sequence lengths; \texttt{H2048} is also slower at 40K and 65K, but reaches comparable throughput at 98K. For these configurations, the memory savings do not consistently offset the execution cost associated with their larger expert-weight volume. Unlike the fixed-batch latency comparison in Section~\ref{sec:eval_moe}, this study decreases the micro-batch as sequence length increases, keeping the number of tokens processed per step approximately constant. Longer sequences therefore do not provide more tokens to amortize expert transfers. Section~\ref{sec:eval_analysis} examines token routing as an alternative when moving experts is costly.

\begin{figure*}[h]
    \centering
    \includegraphics[width=0.95\linewidth]{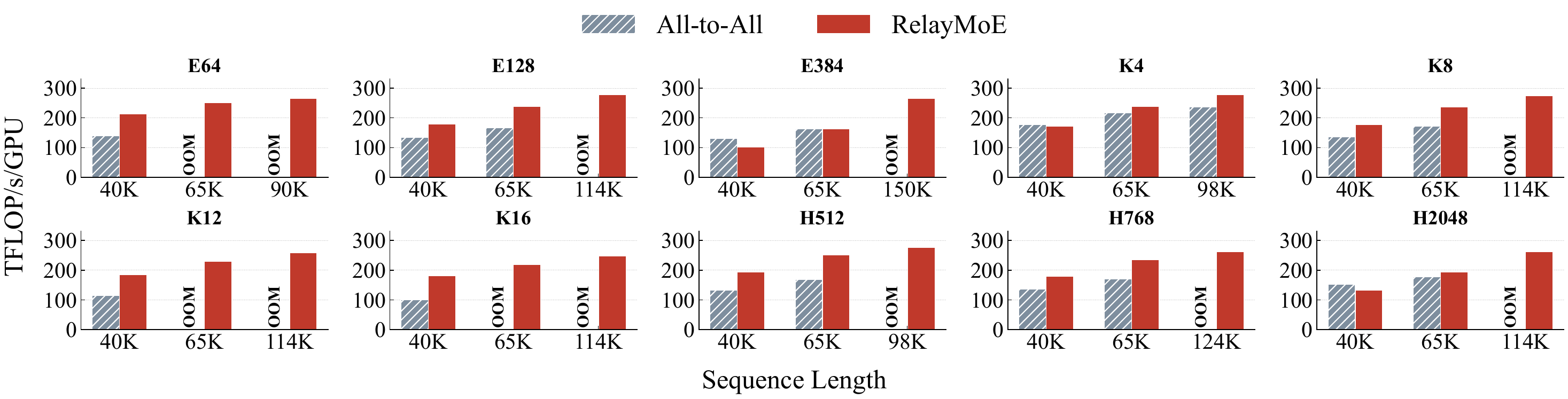}
    \vspace{-10pt}
    \caption{Training throughput as sequence length increases, with micro-batch size $b=12$ and full activation recomputation.}
    \vspace{-5pt}
    \label{fig:eval_seq}
\end{figure*}

\noindent\textbf{Batch scaling for improved throughput.}
We evaluate larger micro-batches as another use of the saved memory. We use the same 8-GPU setup as in the recomputation study and fix the sequence length to 40{,}960. Figure~\ref{fig:batch_scaling} reports three settings for each configuration: all-to-all at micro-batch size $b=12$, \name at the same micro-batch size, and \name at the largest micro-batch size that fits in GPU memory.

At $b=12$, \name is faster in most configurations under the selected checkpoint schedules and reduces full-model peak GPU memory by up to 35.7\% relative to all-to-all. The saved memory allows micro-batches of 18--32, depending on the model. For \texttt{H2048}, \name is slower than all-to-all at $b=12$ but becomes faster at $b=30$; \texttt{E384} also closes most of its throughput gap at $b=28$. Processing more tokens per step can amortize expert transfers, although the attention-checkpoint schedules also change, so these results reflect their combined effect.

The largest batch is not always the fastest: \texttt{E64}, \texttt{K12}, \texttt{K16}, \texttt{H512}, and \texttt{H768} achieve higher throughput with \name at $b=12$. Larger batches consume activation memory that could otherwise retain attention activations, potentially requiring more recomputation. These results show that the memory savings provide a choice between larger batches and less recomputation; batch size and checkpointing need to be considered together to improve throughput.

\noindent\textbf{Sequence scaling for longer context.}
Finally, we evaluate whether \name supports longer sequences when both dispatchers use the same micro-batch size and full recomputation. We use the same 8-GPU setup with $b=12$ and recompute all attention blocks and MoE activations. In Figure~\ref{fig:eval_seq}, only \texttt{K4} trains beyond 65K with all-to-all, whereas \name reaches 90K--150K across configurations, extending the largest tested feasible sequence length by up to 2.85$\times$. This capacity gain persists with attention recomputation held fixed: checkpointing discards forward activations, but backward still needs temporary expert intermediates. \name avoids the full all-to-all dispatch buffers and reconstructs these intermediates within each backward hop, leaving more memory for longer inputs. The tested limits also depend on layer count, retained attention and checkpoint inputs, and model state.

\begin{figure}[!htbp]
    \centering
    \includegraphics[width=\linewidth]{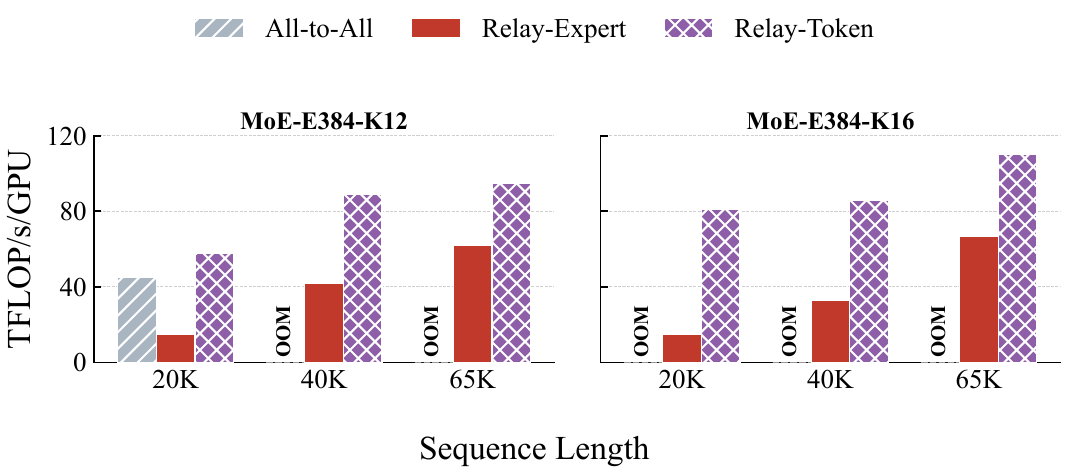}
    \vspace{-15pt}
    \caption{Throughput with token routing, expert routing, and all-to-all on the two 384-expert configurations.}
    \vspace{-10pt}
    \label{fig:adaptive_routing}
\end{figure}
\subsection{Design Analysis}
\label{sec:eval_analysis}

We examine when each routing mode is effective and how communication overlap and GroupedGEMM improve ring execution.

\noindent\textbf{Adaptive routing: token vs expert.}
To evaluate when token routing is preferable, we compare token routing, expert routing, and all-to-all on \texttt{MoE-E384-K12} and \texttt{MoE-E384-K16}, each with 384 experts, hidden dimension and expert FFN width 2048, and 19.53B parameters. All three methods use the same micro-batch size and recomputation settings at each sequence length.

Figure~\ref{fig:adaptive_routing} shows that token routing outperforms expert routing on both configurations. On \texttt{MoE-E384-K12}, its speedup over expert routing decreases from 3.7$\times$ at 20K to 2.1$\times$ at 40K. Token routing also outperforms all-to-all at 20K and runs at 40K and 65K, where all-to-all encounters OOM. On \texttt{MoE-E384-K16}, both ring modes run at all tested sequence lengths while all-to-all encounters OOM. Token routing therefore retains the capacity benefit in these wide-expert configurations while avoiding the cost of circulating their weights.

The narrowing gap between the two ring modes follows the same volume trade-off as the crossover in Section~\ref{sec:eval_moe}: token communication grows with sequence length, while expert-weight communication remains fixed. Together, these results show how token routing complements expert routing when expert-weight volume is large, motivating the routing choice in Section~\ref{sec:adaptive}.

\begin{figure}[!htbp]
    \centering
    \includegraphics[width=\linewidth]{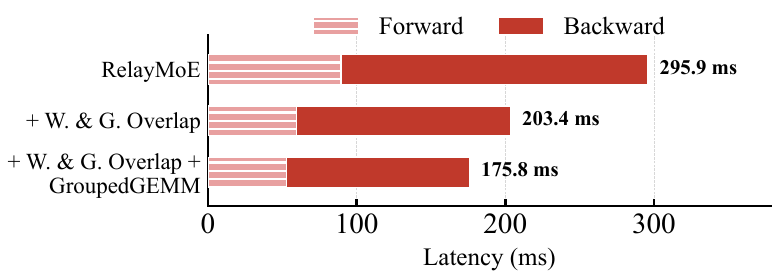}
    \vspace{-15pt}
    \caption{Forward and backward latency with communication--computation overlap and GroupedGEMM enabled successively. ``W.\&G. overlap'' enables weight and gradient transfers to overlap with computation.}
    \vspace{-10pt}
    \label{fig:overlap_ablation}
\end{figure}

\noindent\textbf{Communication overlap and GroupedGEMM.}
To quantify the contribution of \name's optimizations, we measure forward and backward pass time for a single \texttt{E128} MoE layer with EP=8 and fixed token counts. In Figure~\ref{fig:overlap_ablation}, the naive ring implementation executes communication and computation sequentially, without GroupedGEMM, taking 295.9\,ms per iteration. With communication--computation overlap in the second bar, total latency decreases to 203.4\,ms: transfers proceed while computation is ongoing at each hop, reducing exposed communication time. Adding GroupedGEMM in both forward and backward further reduces latency to 175.8\,ms by processing the current partition's expert computations together. The additional reduction shows that improving local expert computation complements communication overlap within the ring.

\begin{figure}[!htbp]
    \centering
    \includegraphics[width=\linewidth,trim=4bp 14bp 12bp 12bp,clip]{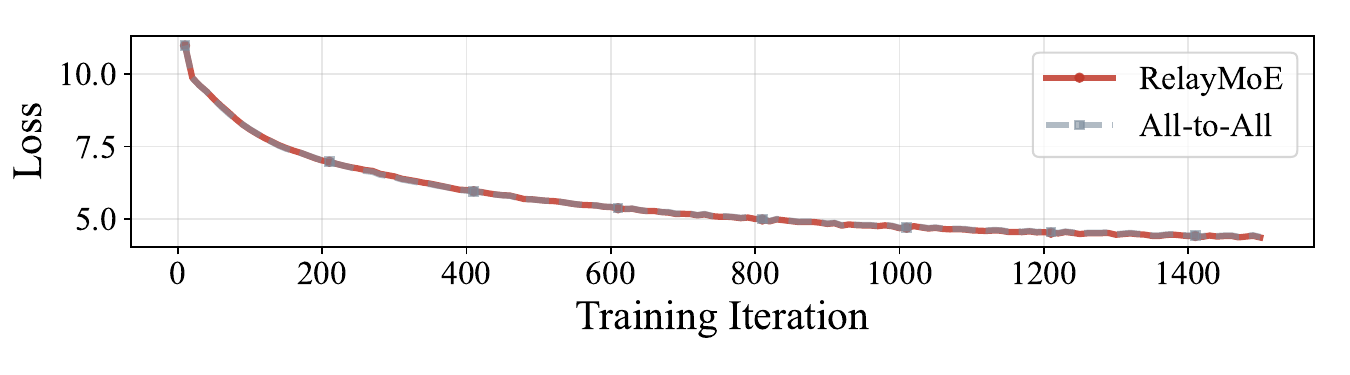}
    \vspace{-15pt}
    \caption{Training loss of Qwen3-30B-A3B with \name and all-to-all, using the same initialization and hyperparameters.}
    \vspace{-10pt}
    \label{fig:loss}
\end{figure}


\noindent\textbf{Training correctness}. We verify that \name preserves training correctness relative to the standard all-to-all dispatcher. We train Qwen3-30B-A3B with \name and all-to-all under the same training settings, including initialization, random seed, and hyperparameters. Figure~\ref{fig:loss} shows that the two loss curves are indistinguishable over 1500 training iterations. This is consistent with the mathematical equivalence of the two formulations (Section~\ref{sec:local}): both compute the same token--expert pairs, with possible floating-point rounding differences from the accumulation order.

\section{Related Work}
\label{sec:related_work}

\noindent\textbf{MoE training}. Existing work optimizes communication in MoE expert parallelism through dispatch libraries~\cite{deepep2025,hybridep2025,ncclep,ucclep2025}, communication--computation overlap~\cite{comet2025,tutel,lina2023,schemoe2024,fsmoe2025,deepseek2024v3,lancet2024}, and network design~\cite{mixnet2025}. Other work optimizes expert computation and activation memory usage~\cite{megablocks2023,moeblaze2026,sonicmoe2025}, parallelism and pipeline layouts~\cite{megascale2026,moeparallelfolding2025,mpipemoe2023,foldmoe2025}, and execution on HPC platforms~\cite{xmoe2025}. \name targets the top-$k$-expanded inter-GPU dispatch buffers. Its ring execution can incorporate optimized expert kernels and communication--computation overlap, and complements attention parallelism.

To reduce communication overhead, existing work explores transferring experts to GPUs that host tokens. FasterMoE~\cite{fastermoe2022} replicates overloaded experts near their tokens to alleviate load imbalance. In its data-centric mode, Janus~\cite{janus2023} uses credits to bound GPU expert-fetch buffers and caches fetched weights in host memory for backward. \name targets the peak from expert intermediates during backward, reconstructing and releasing them within each hop using the same weight transfers as gradient computation. These memory savings support longer sequences, larger batches, and larger top-$k$ values within the same GPU memory budget.

MemFine~\cite{memfine2025} reduces MoE activation memory usage through token chunking and chunk-level recomputation, retaining dispatch--compute--combine for each chunk. Ring-Expert in \name likewise limits the lifetime of reconstructed intermediates, but keeps tokens local and circulates expert partitions, avoiding inter-GPU token-dispatch buffers while reusing weight transfers for recomputation and gradient computation.


\label{sec:related-mem}

\noindent\textbf{Memory-efficient training}. Offloading~\cite{zerooffload2021,zeroinfinity2021}, model-state partitioning~\cite{zero}, and activation checkpointing~\cite{chen2016,korthikanti2023reducing} have been explored to reduce memory pressure in LLM training.
These techniques reduce GPU memory usage for model states or trade recomputation for fewer retained activations.
\name complements them by reducing the memory required during MoE execution, including the intermediates reconstructed during backward.
The memory saved in MoE layers can be used to retain more attention activations, reducing attention recomputation and improving throughput for the same training workload.

Other work reduces the memory cost of attention. FlashAttention~\cite{flashattn2022,flashattn2_2024,flashattn3_2024} reduces attention activation memory usage, while sequence and context parallel systems distribute attention computation and activations across GPUs~\cite{seqpar2023,megatron-lm,ulysses2023,ringattention2024,fang2024uspunifiedsequenceparallelism,gu2024loongtrainefficienttraininglongsequence,ringx2025}.
RingAttention also uses a ring, but circulates KV blocks for attention; \name addresses MoE dispatch buffers and expert activations.
Reducing attention memory leaves these MoE costs in place, so the MoE sublayer can still limit the trainable sequence length.
These attention optimizations and \name can be combined to reduce memory in both sublayers, supporting longer-context training under the same hardware budget.

\section{Conclusion}

We presented \name, a ring-based MoE execution model that avoids full top-$k$-expanded dispatch buffers and reduces the memory required for expert recomputation during backward. The saved memory supports longer sequences and larger micro-batches, and can be used to reduce attention recomputation. Our evaluation demonstrates lower peak GPU memory usage and improved training throughput over all-to-all dispatch.


\bibliographystyle{plain}
\bibliography{reference}

\end{document}